\documentclass[conference]{IEEEtran}
\IEEEoverridecommandlockouts

\def\BibTeX{{\rm B\kern-.05em{\sc i\kern-.025em b}\kern-.08em
    T\kern-.1667em\lower.7ex\hbox{E}\kern-.125emX}}

\usepackage{comment}
\usepackage[noadjust]{cite}
\usepackage{amsmath,amssymb,amsfonts}
\usepackage{listings}
\usepackage{graphicx}
\usepackage{textcomp}
\usepackage[table]{xcolor}
\usepackage{ragged2e}
\usepackage{enumitem}
\usepackage{bbm}
\usepackage{placeins}
\usepackage{comment}
\usepackage{pgfgantt}

\usepackage{lipsum}
\usepackage{multicol}
\usepackage{url}
\usepackage[linesnumbered,ruled]{algorithm2e}
\usepackage{algpseudocode}

\usepackage{tikz}
\usetikzlibrary{shapes.geometric,arrows.meta,positioning,decorations.markings,decorations.pathreplacing}
\tikzstyle{my right of} = [right=of #1.east]
\tikzstyle{my left of} = [left=of #1.west]

\newcommand\copyrighttext{%
  \footnotesize \textbf{To appear in IEEE AITesting 2021} \\
  \footnotesize \textcopyright 2021 IEEE. Personal use of this material is permitted. Permission from IEEE must be obtained for all other uses, in any current or future media, including reprinting/republishing this material for advertising or promotional purposes, creating new collective works, for resale or redistribution to servers or lists, or reuse of any copyrighted component of this work in other works.}
\newcommand\copyrightnotice{%
\begin{tikzpicture}[remember picture,overlay]
\node[anchor=south,yshift=15pt] at (current page.south) {\fbox{\parbox{\dimexpr\textwidth-\fboxsep-\fboxrule\relax}{\copyrighttext}}};
\end{tikzpicture}%
}

\usepackage{subcaption}

\begin{document}
\title{SilGAN: Generating driving maneuvers for scenario-based software-in-the-loop testing}

\author{
\IEEEauthorblockN{Dhasarathy Parthasarathy\IEEEauthorrefmark{2}\IEEEauthorrefmark{1}, Anton Johansson\IEEEauthorrefmark{1}} \\
\IEEEauthorblockA{\IEEEauthorrefmark{2}Volvo Group, Gothenburg, Sweden, Email: dhasarathy.parthasarathy@volvo.com}
\IEEEauthorblockA{\IEEEauthorrefmark{1}Chalmers University of Technology, Gothenburg, Sweden, Email: johaant@chalmers.se}
}

\maketitle
\copyrightnotice

\begin{abstract}
Automotive software testing continues to rely largely upon expensive field tests to ensure quality because alternatives like simulation-based testing are relatively immature. As a step towards lowering reliance on field tests, we present SilGAN, a deep generative model that eases specification, stimulus generation, and automation of automotive software-in-the-loop testing. The model is trained using data recorded from vehicles in the field. Upon training, the model uses a concise specification for a driving scenario to generate realistic vehicle state transitions that can occur during such a scenario. Such authentic emulation of internal vehicle behavior can be used for rapid, systematic and inexpensive testing of vehicle control software. In addition, by presenting a targeted method for searching through the information learned by the model, we show how a test objective like code coverage can be automated. The data driven end-to-end testing pipeline that we present vastly expands the scope and credibility of automotive simulation-based testing. This reduces time to market while helping maintain required standards of quality. \looseness=-1
\end{abstract}

\lstset{
   basicstyle=\fontsize{7}{10}\selectfont\ttfamily,
   commentstyle=\color{white},
   emph={lt},emphstyle={\color{black}\bfseries}
}
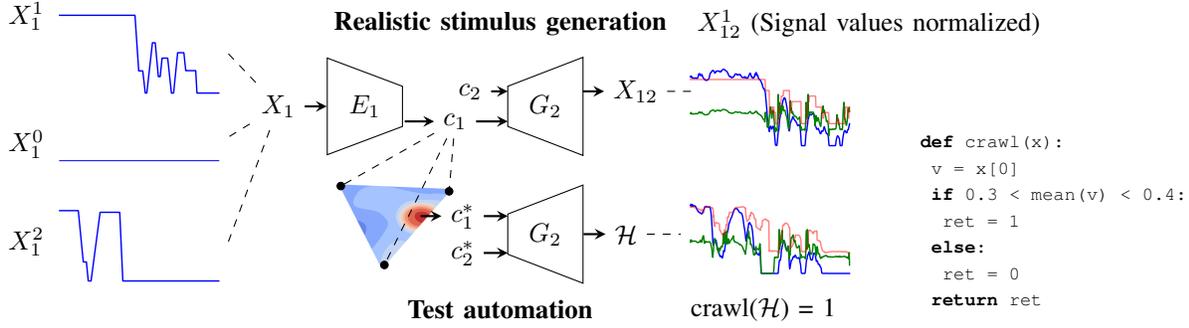
\begin{figure*}[!ht]
    \centering
    
\begin{minipage}[b]{.65\linewidth}
        \begin{tikzpicture} [transform shape]

\tikzstyle{trapez} = [draw, trapezium, trapezium stretches=true, minimum height=1.0cm,
  text width=0.3cm, align=center]
  
\tikzstyle{arrow} = [thick,->,>=stealth]
  
\node (Encoder1) at (0,0) [trapez, rotate=270] {\rotatebox{-270}{$E_1$}};

\node (Decoder2) at (2.4,0) [trapez, rotate=90] {\rotatebox{-90}{$G_2$}};

\draw node[left=0.8cm of Encoder1.center] (Template) {$X_1$};
\draw [arrow] (Template) -- (Encoder1);

\draw node[] at (1.2,-0.2) (contentCode1) {$c_{1}$};
\draw [arrow] (0.53,-0.2) -- (contentCode1);
\draw [arrow] (contentCode1) -- (1.9, -0.2);

\draw node[] at (1.4,0.2) (contentCode2) {$c_2$};
\draw [arrow] (contentCode2) -- (1.9, 0.2);

\draw node[] at (3.6,0.2) (translatedSignal) {$X_{12}$};
\draw [arrow] (2.9,0.2) -- (translatedSignal);

\draw[line width=1.pt, smooth,samples=100,domain=0.7:1.2] plot[parametric] function{t,2.718281828459045**(-(t-0.95)**(2.0)*80.0)/3.7+0.10};

\newcommand\imgWidth{0.2}
\newcommand\imgWidthDouble{0.33}

\node[inner sep=0pt] (T0) at (-3,-0.2)
    {\includegraphics[width=\imgWidth\textwidth]{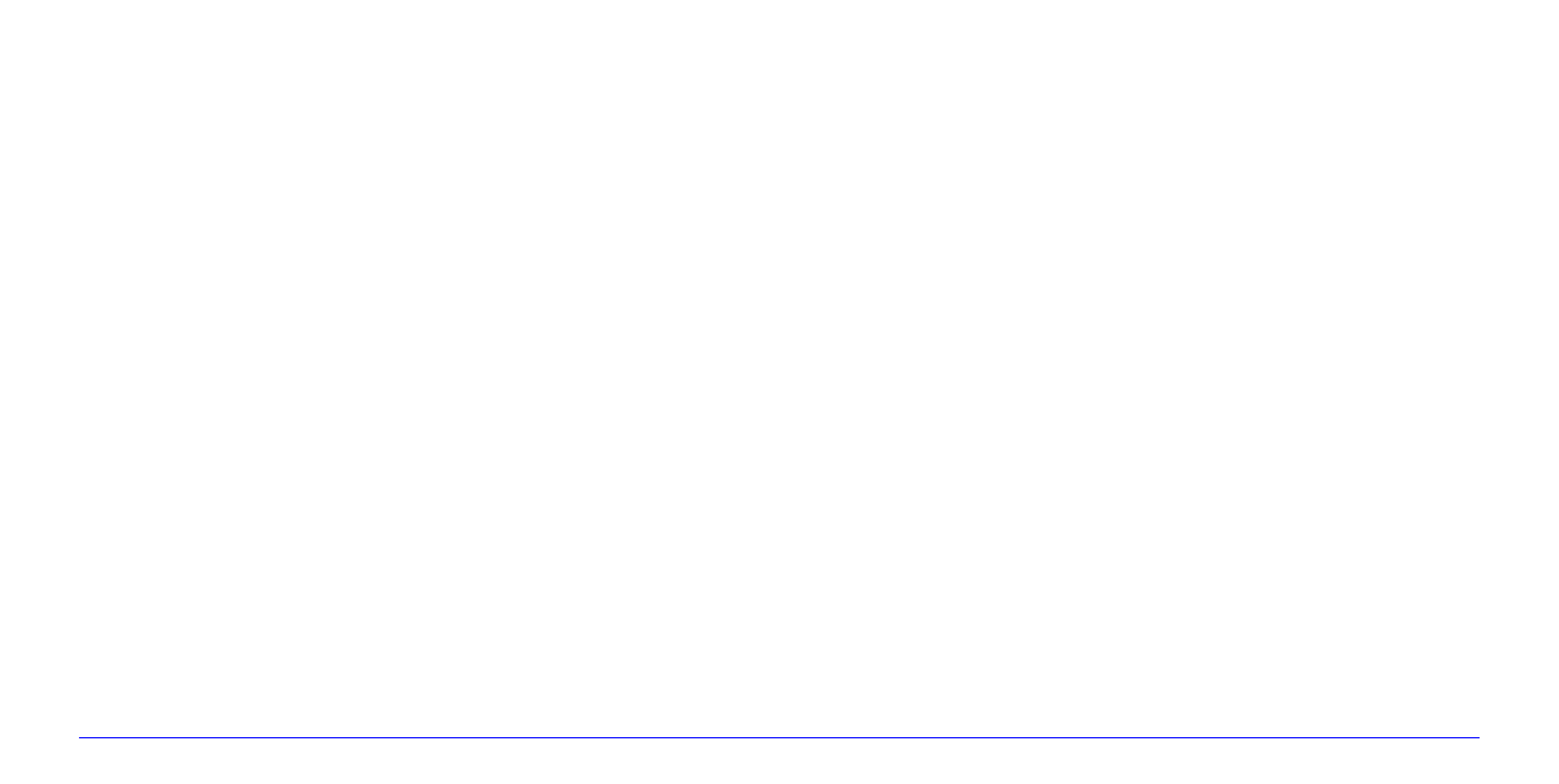}};
    
\node[inner sep=0pt] (T1) at (-3,0.7)
    {\includegraphics[width=\imgWidth\textwidth]{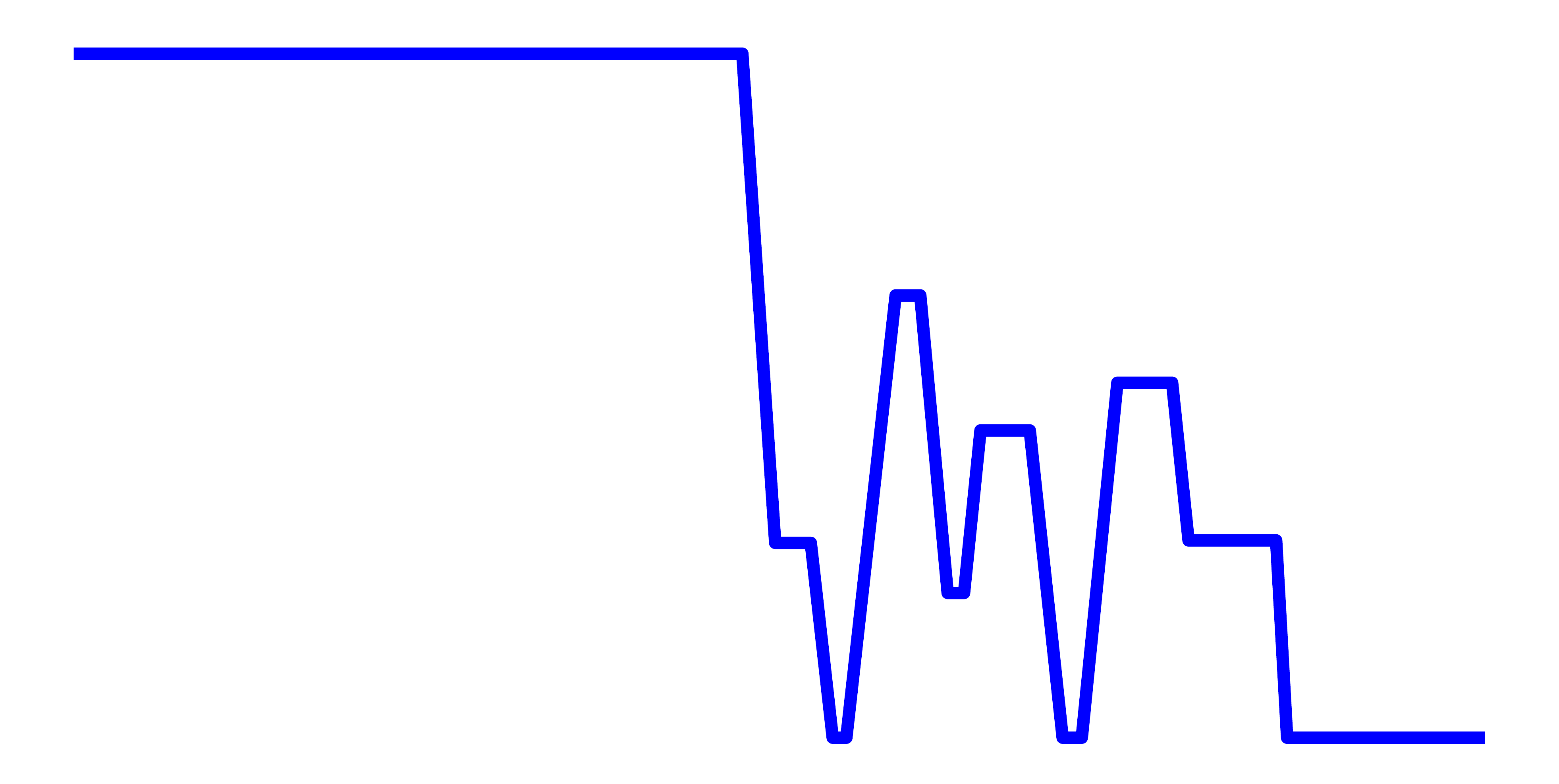}};

\node[inner sep=0pt] (T2) at (-3,-1.8)
    {\includegraphics[width=\imgWidth\textwidth]{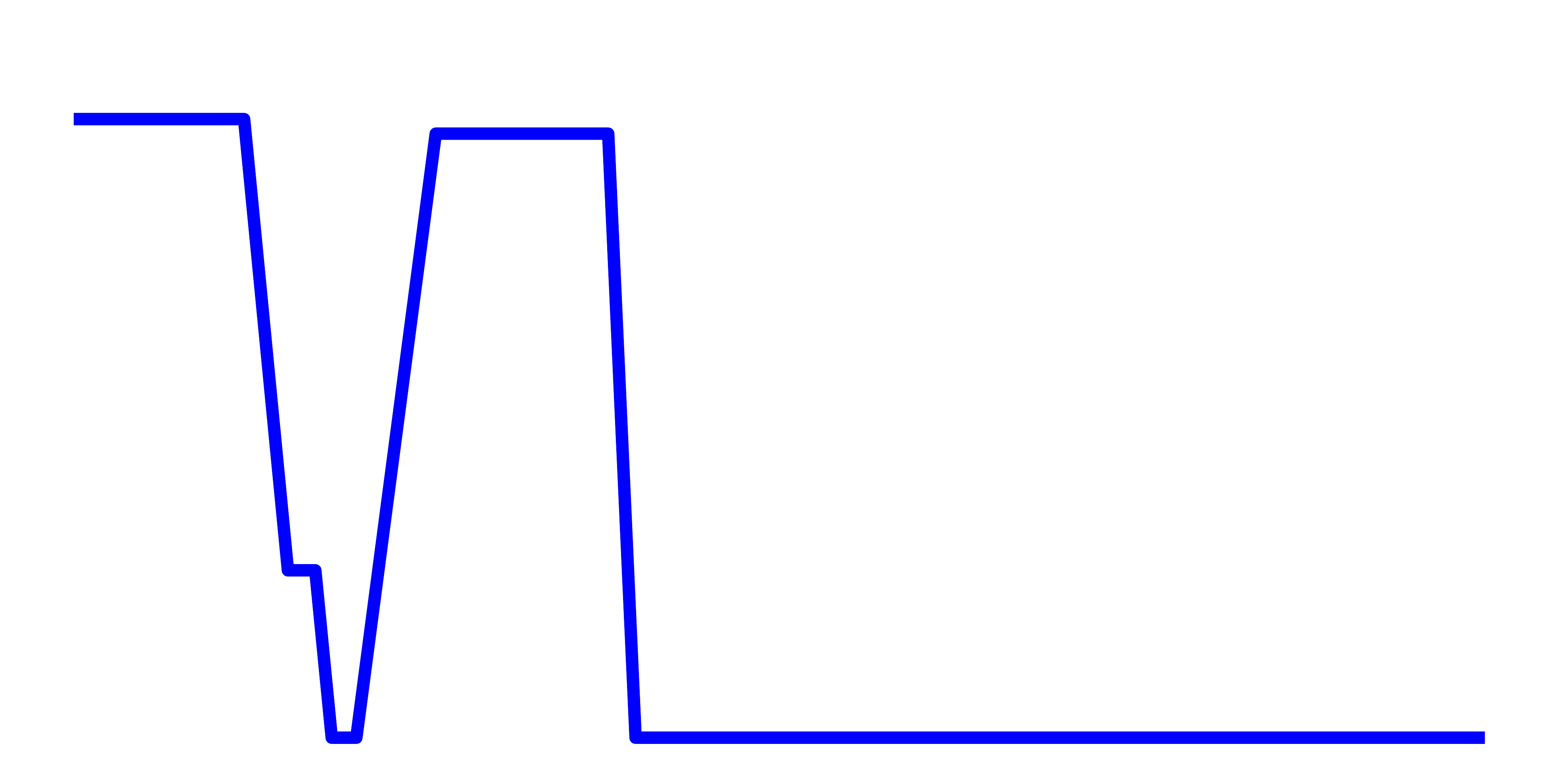}};

\draw [dashed] (-1.8,-0.4) -- (Template);
\draw [dashed] (T1.east) -- (Template);
\draw [dashed] (T2.east) -- (Template);

\node[inner sep=0pt] (T2) at (0.4,-1.6)
    {\includegraphics[width=0.13\textwidth]{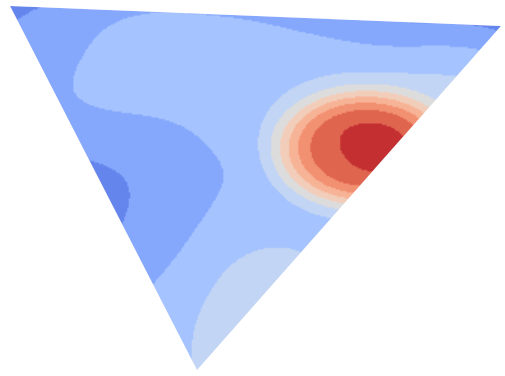}};

\filldraw [black] (1.12,-1.13) circle (0.05cm);
\filldraw [black] (-0.32,-1.06) circle (0.05cm);
\filldraw [black] (0.26,-2.11) circle (0.05cm);

\draw [dashed] (contentCode1) -- (1.12,-1.13);
\draw [dashed] (contentCode1) -- (-0.32,-1.06);
\draw [dashed] (contentCode1) -- (0.26,-2.11);

\draw node[] at (1.3,-1.46) (gradContentCode) {$c_1^*$};
\draw node[] at (1.3,-1.95) (gradStyleCode) {$c_2^*$};
\draw [arrow] (0.74,-1.46) -- (gradContentCode);

\node (Decoder2_bottom) at (2.4,-1.7) [trapez, rotate=90] {\rotatebox{-90}{$G_2$}};

\draw [arrow] (gradContentCode) -- (1.9, -1.46);
\draw [arrow] (gradStyleCode) -- (1.9, -1.95);
\draw node[right=0.8cm of Decoder2_bottom.center] (validSignal) {$\mathcal{H}$};
\draw [arrow] (Decoder2_bottom) -- (validSignal);

\node[right=1.8cm of Decoder2_bottom.center, inner sep=0pt] (Hit)
    {\includegraphics[width=\imgWidth\textwidth]{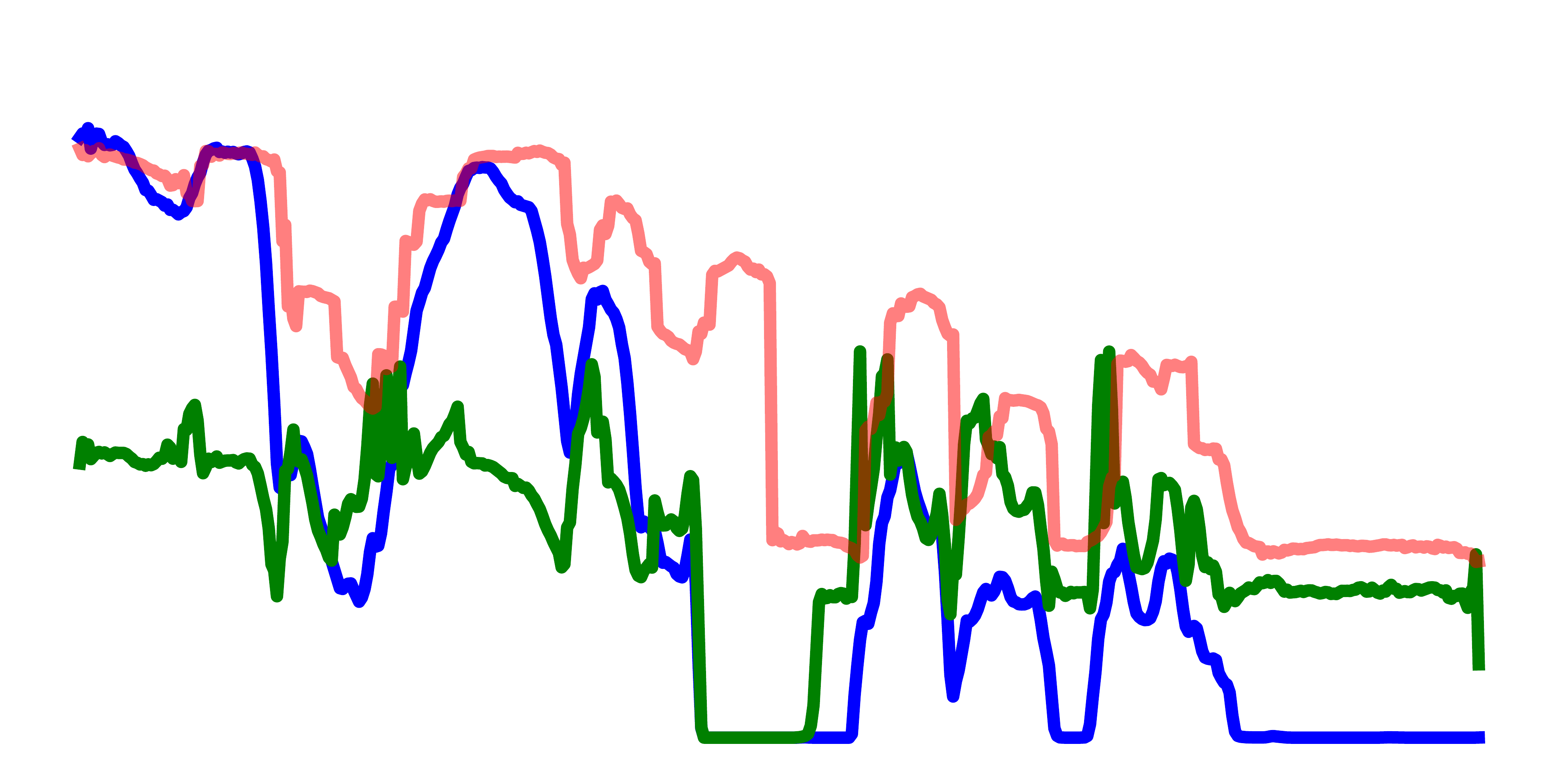}};
\draw [dashed] (validSignal) -- (Hit);

\node[right=1.8cm of Decoder2.center, inner sep=0pt] (translatedImg)
    {\includegraphics[width=\imgWidth\textwidth]{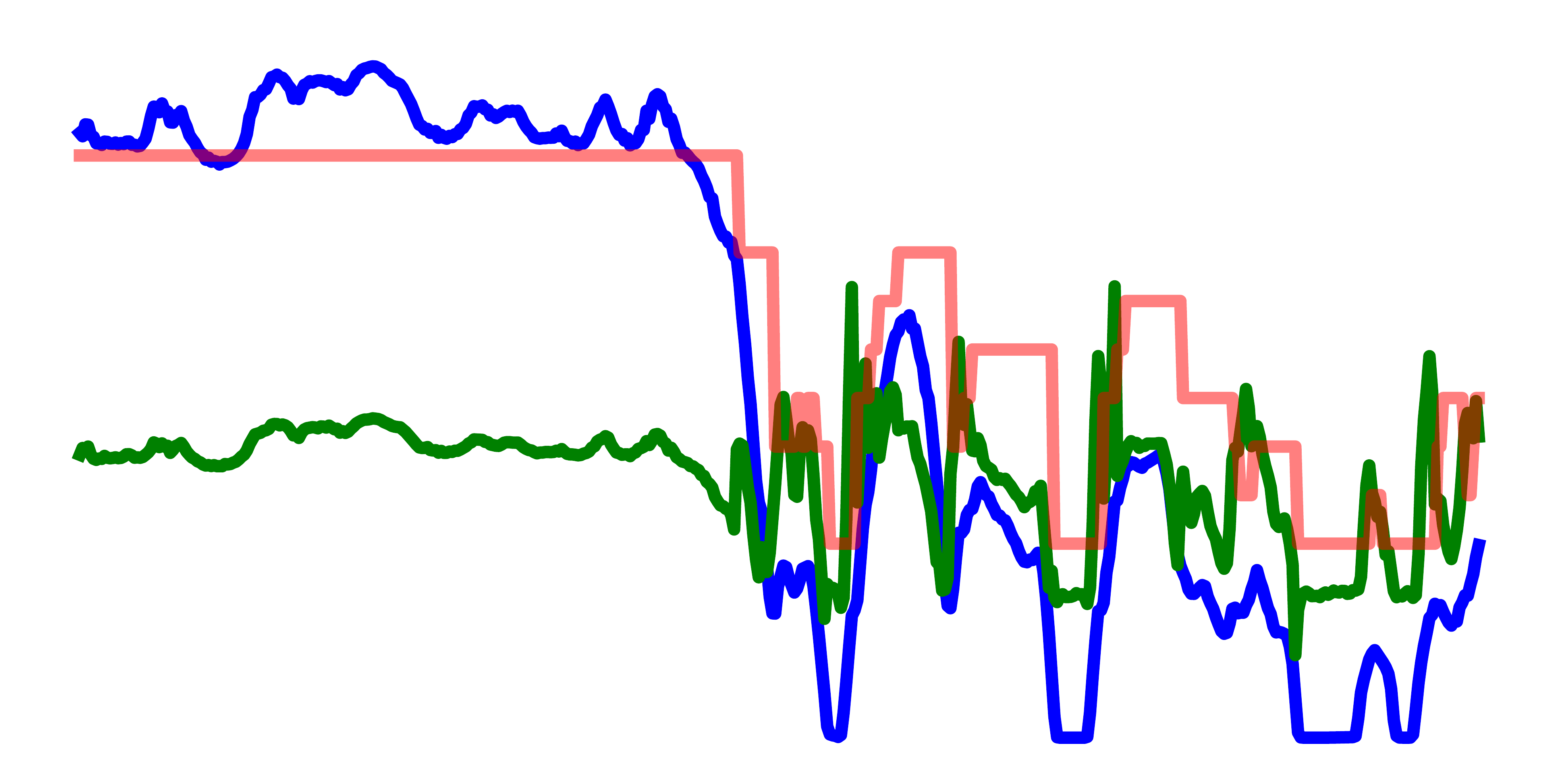}};

\draw [dashed] (translatedSignal) -- (4.4,0.2);

\draw node[] at (-3.0, 1.7) {\textbf{Test scenario specification}};
\draw node[] at (-4.5, 1.2) {\textbf{$X^1_1$}};
\draw node[] at (-4.5, -0.5) {\textbf{$X^0_1$}};
\draw node[] at (-4.5, -1.8) {\textbf{$X^2_1$}};

\draw node[] at (1.8, 1.1) {\textbf{Realistic stimulus generation}};
\draw node[] at (6.7, 1.1) {\textbf{$X^1_{12}$} (Signal values normalized)};

\draw node[] at (1.8, -2.7) {\textbf{Test automation}};
\draw node[] at (5.3, -2.7) {crawl($\mathcal{H}$) = 1};

\end{tikzpicture}
    \end{minipage}\qquad
    \begin{minipage}[b]{.25\linewidth}
       
\begin{lstlisting}[language=Python]
def crawl(x):
 v = x[0]
 if 0.3 < mean(v) < 0.4:
  ret = 1
 else:
  ret = 0
 return ret
\end{lstlisting}    
    \end{minipage}

    \caption{Overview of SilGAN - A test scenario (Section \ref{sec:templ}) of a vehicle coming to a stop can be specified using template $X^1_1$, which is a rough sketch of how the vehicle speed is expected to vary. We train SilGAN (Section \ref{sec:model}) to translate this template into a maneuver $X^1_{12}$, containing a realistic rendition of the template along with related transitions of engine speed and selected gear. Such a generated realistic maneuver can be applied as stimuli for software under test. The test scenario can also be arbitrarily enriched by including conditions like template $X^2_1$, depicting another variation of a vehicle stopping. Generated maneuvers can thus be constrained to combine only the characteristics of specified templates (Section \ref{sec:gen}). Having defined such a composite scenario, we show how SilGAN can be used for test automation (Section \ref{sec:aut}). Consider $crawl$, a simple Python function that checks whether a vehicle is crawling at a relatively low speed. Using feedback from a transformed version of $crawl$, we demonstrate a method that searches through the model's latent information spaces to automatically generate a realistic stop maneuver $\mathcal{H}$. Generated test stimulus $\mathcal{H}$ not only complies with the specified test scenario, but also makes $crawl(\mathcal{H})=1$.\looseness=-1}
    \label{fig:model_overview}
    \vspace{-5mm}
\end{figure*}
\section{Introduction}
\label{sec:intro}
Embedded software lies at the heart of a variety of internal vehicle functions such as engine control, driver assistance, and energy management \cite{DBLP:journals/jss/HaghighatkhahBP17}. As modern software-driven commercial vehicles grow more capable, they are becoming indispensable parts of a variety of industrial activities such as urban distribution, public transport, construction, and mining. This, however, creates a growing demand for new functionality, placing increased pressure on vehicle manufacturers to rapidly develop, test, and deploy new software. Increasing the pace of development, while continuing to deliver quality software, are urgent priorities that the automotive industry faces today \cite{DBLP:journals/smr/StolfaSBMDBM20}. One way in which the industry has responded is by taking a platform approach. Today, it is typical for core applications like driveline or battery management to be re-used across different vehicle platforms. While this streamlines engineering effort, the applications themselves begin to be used in a variety of different ways. This inevitably increases the chance that they regress when being used in poorly understood or unanticipated ways. For example, electrically steered axles of a truck could draw higher than expected current only when it rolls downhill at an inclination and surface friction that is common in pit mines. In trying to identify and avoid such failures, vehicle manufacturers typically spend significant effort, subjecting software to thousands of kilometers worth of time consuming field tests. \looseness=-1

Recognizing the clear need to accelerate test feedback, the industry has been turning more towards simulation-based testing with software-in-the-loop (SIL) \cite{kaijser:hal-02156371}. The vehicle software system consists of Software Components (SWCs), distributed over electronic control units, which exchange information using \textit{signals}. Each signal captures one element of vehicle state, such as the current speed or steering angle. SWCs realize their functions by responding to transitions in signals which are routed to them. Therefore, for credible simulation-based testing with SIL, especially for low-level control software, test setups need to generate signal transitions at a deep level of detail. Today's SIL tests rigs simulate these transitions by manually specifying explicit, domain specific, mathematical rules in the form of plant models. This works well when testing individual SWCs, where plant models need to simulate transitions of only a handful of signals. As the scope expands to testing larger sub-systems of SWCs, under the influence of realistic driving, modeling takes significant effort. This is because, during a driving maneuver that lasts several minutes, the continuous, multi-dimensional vehicle state-space undergoes complex transitions, many of which are difficult to model. \looseness=-1

However, the ongoing data revolution in transportation means that vehicle manufacturers now have unprecedented insights into how their vehicles are being used in the field. The availability of internal vehicle data, combined with rapid advances in deep learning, makes it possible to learn intricate state transitions that occur during different driving scenarios. Leveraging techniques of deep generative modeling, in this work, we contribute the following (refer Figure \ref{fig:model_overview}).

\begin{itemize}[leftmargin=*]
    \item We introduce \textit{templates} as a simple, concise and granular specification of a driving scenario
    \item We train a model to translate a template into realistic multi-dimensional signal transitions. Transitions, so generated, emulate authentic vehicle behavior during the specified scenario and can be used as stimulus for software under test. \looseness=-1
    \item For given software under test, we show how stimulus generation can be automated to satisfy a specific test objective,  \looseness=-1
\end{itemize}

This novel set of contributions significantly expand the credibility of simulation-driven automotive software testing, further lowering the reliance on field tests. This reduces time and cost to market, while helping maintain the necessary standards of quality. Code from this work is publicly available\footnote{https://github.com/dhas/silGAN}.\looseness=-1

\section{Test scenarios for control software}
\label{sec:templ}
As surveyed in \cite{847181e9c5cb4c2e8a0adae5107abda7}, recent years have seen many proposals that help specify scenarios for testing automotive software systems. However, a major limitation that they share is a focus on specifying scenarios only in terms of high-level driving characteristics. For example, popular ontologies like \cite{DBLP:conf/itsc/UlbrichMRSM15} describe driving scenarios mostly in terms of tactical aspects like avoiding obstacles or changing lanes. Such proposals may have been used to analyze and test driver assistance or autonomous driving controllers (for example, \cite{DBLP:conf/itsc/FremontKPSABWLL20}), but they define scenarios at too low a level of granularity. Vehicle control software at the operational level, i.e. those for functions like fuel-injection or automatic gear shifting, react to transitions of a multitude of internal vehicle signals. Testing these controllers under different driving scenarios would therefore require realistic specification of these signals at a granular level of detail. The current practice of manual plant modeling addresses this to a certain extent. But, when the scope of testing widens or the time-horizon of testing extends beyond a few seconds, plant models need support in capturing complex, non-linear and unforeseen internal vehicle phenomena.\looseness=-1

\subsection{Maneuver - a multi-dimensional time series of signals}
Vehicles often execute recognizable driving patterns, and we refer to a family of similar patterns as a driving \textit{scenario}. One example of a driving scenario would be takeoff, where the vehicle starts rolling and subsequently begins to cruise. Depending upon a variety of factors like vehicle mass, road inclination, surface and traffic conditions, there are several ways of executing takeoff. We refer to one specific execution of a driving scenario as a driving \textit{maneuver}. Figure \ref{fig:recorded-maneuvers} captures $\sim$10-minute long maneuvers of takeoff and cruise scenarios using three recorded signals vehicle speed, engine speed and selected gear. Takeoff, for instance, is interesting because during its execution, apart from several core control actions, additional ones like automatic engine start, cabin heat circulation, and parking brake release are activated. Testing the behavior of such functions under many possible instances of takeoff can therefore be of great value. However, as noted earlier, if one were to collectively test SWCs for all these controllers, then detailed signal transitions on the form shown in Figure \ref{fig:recorded-maneuvers} must be applied as test input. With numerous possible variations of takeoff, manually specifying transitions of several signals, for even a handful of takeoff maneuvers, is clearly difficult.\looseness=-1

\subsection{Template - a 1-D signal-level scenario description}
\begin{figure}[t]
    \begin{subfigure}{\linewidth}
        \centering
        \includegraphics[width=0.9\linewidth]{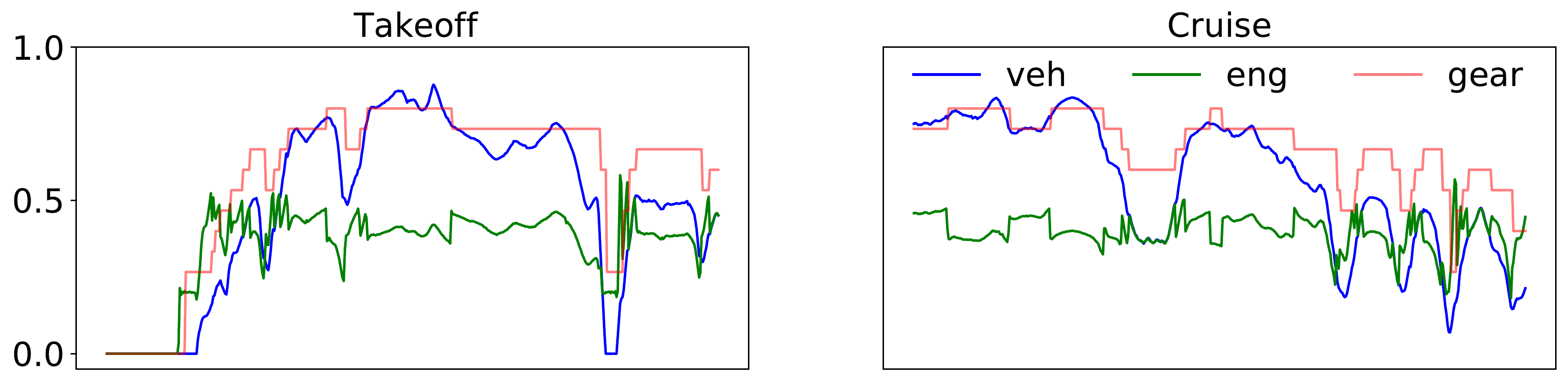}
        \caption{}
        \label{fig:recorded-maneuvers}
    \end{subfigure}
    \begin{subfigure}{\linewidth}
        \centering
        \includegraphics[width=0.9\linewidth]{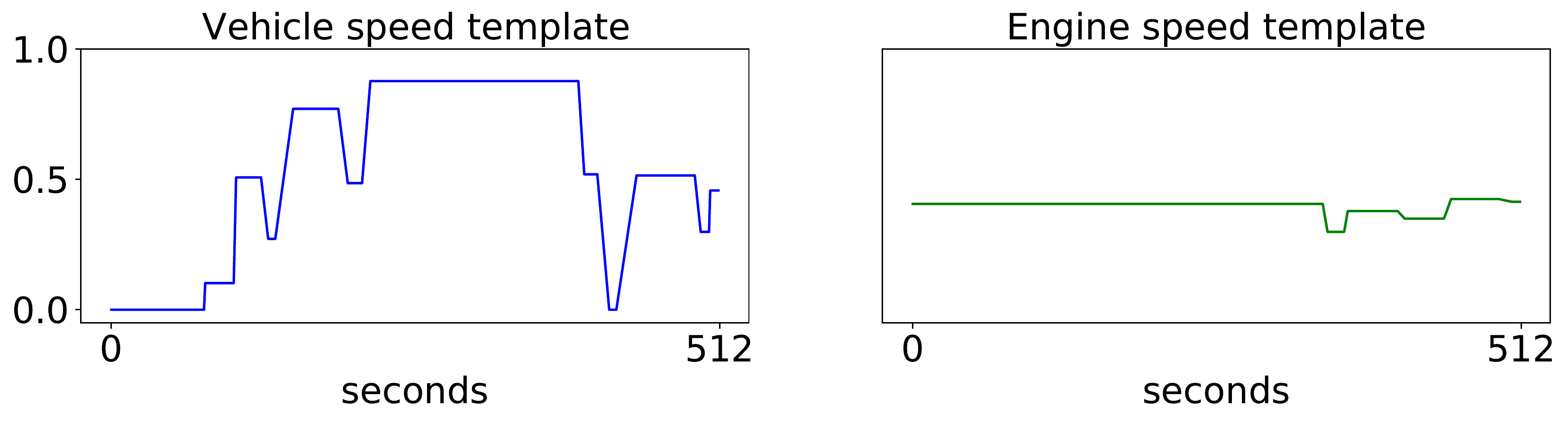}
        \caption{}
        \label{fig:templates}
    \end{subfigure}
    \caption{Recorded driving maneuvers (a) and corresponding templates (b). Signal values normalized}
    \vspace{-5mm}
    \label{fig:maneuvers-templates}
    \setlength{\belowcaptionskip}{-10pt}
\end{figure}

The difficulty in specifying a realistic test maneuver arises from two problems - (i) the need to specify detailed short and long term transitions of each constituent signal, and (ii) capturing possible inter-signal dependencies. What would help, therefore, is a specification mechanism that ignores details and focuses only on the basic profile of the maneuver. This can be followed by another mechanism that can translate this profile into a realistic maneuver. Building upon previous work for test stimulus generation \cite{DBLP:conf/aitest/ParthasarathyBH20}, we propose piece-wise linear approximation as a simple way to avoid specifying transient details of a signal. The resulting \textit{template}, as shown in Figure \ref{fig:templates}, serves as a rough, intuitive sketch of the long-term signal profile. To address the specification of inter-signal dependencies, one could use the proposal in \cite{DBLP:conf/aitest/ParthasarathyBH20} to jointly specify a template for each signal included in the driving maneuver. In practice, however, specifying plausibly related templates for all signals is quite difficult. We therefore propose a much simpler approach of specifying a driving scenario by sketching a template for \textit{any one} constituent signal. This completely eliminates the need to address dependencies. As seen in Figure \ref{fig:templates}, this means that a tester can choose to define a takeoff scenario by specifying only a vehicle speed template. The same tester can choose to define a cruise scenario by only specifying a template for the engine speed. This flexibility allows test case design to focus on a single chosen control signal, further reducing the specification effort. By thus specifying long-range driving scenarios at level of detail suitable for testing entire sub-systems of SWCs, templates are also different from alternative proposals like \cite{10.1145/2786805.2804432}, which focus on specifying test scenarios for individual software routines. \looseness=-1

Templates may simplify the specification of a test scenario for low-level control software, but translating them into realistic maneuvers is not straightforward. Not only should such translation restore the details of the sketched signal, but it should also generate realistically inter-related accompanying signals. To achieve this, we train a deep generative model.\looseness=-1

\section{SilGAN - Translating templates to maneuvers}
\label{sec:model}
Let us define the domain of one-dimensional templates as $\mathcal{X}_1$, and that of $L$-dimensional driving maneuvers as $\mathcal{X}_2$. Let $P_l(X_1)$ denote the distribution of templates for the $l\textsuperscript{th}$ signal of the $L$-dimensional system, and $P(X_2)$ denote the distribution of maneuvers. Denoting the duration of templates and maneuvers by $N$, a template $X_1 \sim P_l(X_1)$, is a time-series in $\mathbb{R}^{1 \times N}$, while a maneuver $X_2 \sim P(X_2)$ is a time-series in $\mathbb{R}^{L \times N}$. The objective is to learn $\mathcal{T}: \mathcal{X}_1 \rightarrow \mathcal{X}_2$ that can map $X_1$ into a realistic maneuver $X_{12} \sim P_l(X_2|X_1)$, that is faithful to the specified profile in the $l\textsuperscript{th}$ signal and generates plausible related variations for all $(L-1)$ other  signals. Deep learning techniques that achieve such mappings by implicitly estimating the conditional distribution fall under the general category of deep domain adaptation \cite{DBLP:journals/tist/WilsonC20}. \looseness=-1

\subsection{Training data}
The number of signals included in a driving maneuver clearly plays an important role in defining its scope for testing. Here, to sufficiently showcase the capabilities of the model, we fix $L=3$ and $N=512$. The training data comes from a source of signals recorded at 1 $Hz$ from 19 Volvo buses operating over a 3-5 year period. From this around 200$k$, 512-second long snapshots of three important signals related to vehicle dynamics -- vehicle speed, engine speed, and selected gear -- have been collected for training. Selected signals exhibit a good mix of levels of granularity. Vehicle speed is a high-level characteristic, while engine speed and (automatically) selected gear are detailed characteristics of the driveline. Moreover, these signals are routed to a wide variety of SWCs, making them relevant for a number of SIL test cases. Translating a template into a maneuver may be complex but, given a recorded maneuver $X_2$, piecewise linear templates $X^{(L)}_1 = (X^1_1, ...X^{L}_1)^T$ for all $L$ signals can be automatically extracted using relatively simple procedures. Upon smoothing a signal using windowed mean and 1-D Sobel filtering, its template is approximated as flat regions around major points inflection, with straight-line edges connecting these regions (refer Figure \ref{fig:maneuvers-templates}). The resulting training data therefore consists of pairs $(X^{(L)}_1, X_2)$ of templates $X^l_1 \sim P_l(X^l_1)$ and recorded maneuvers $X_2 \sim P(X_2)$ they were extracted from. In using this training set, it is important to note our assumption that templates specified by a tester can be plausibly sampled from the distribution of extracted templates $P_l(X^l_1)$. Since a template $X_1^l$ is always associated with $l$, the signal which it approximates, for ease of notation, we cease associating it, or any of its mappings, explicitly with $l$. \looseness=-1

\subsection{Model design}
\noindent \textbf{\textit{Translation}} - A translated maneuver $X_{12}$ is expected to realistically render the template $X_1$ as its $l\textsuperscript{th}$ signal. Since $X_1$ and $X_{12}$ clearly share some characteristics, let us define $C_1$ as a latent intermediate domain that encodes this shared information. Let $E_1 : \mathcal{X}_1 \rightarrow C_1$ be a learnable encoder that maps a template into this domain. However, the translated maneuver $X_{12}$ is also required to render realistic transitions of all other signals. Let us therefore define another latent domain $C_2$ as the source of all this additional information. Allowing the learning process to optimally structure its information, codes from this domain are sampled as $c_2 \sim \mathcal{N}(0,I)$. The translation process can then be completed by defining a learnable generator $G_2 : C_1 \times C_2 \rightarrow \mathcal{X}_{2}$, that produces a maneuver by combining information from both latent domains (\ref{eq:fwd-trans}). In order to ensure that the translated maneuver is realistic, a least-squares discriminator \cite{DBLP:conf/iccv/MaoLXLWS17} $D_2 : \mathcal{X}_2 \rightarrow \mathbb{R}$ is defined. Using ground truth labels $0$ for fake samples and $1$ for real samples, it learns to estimate the conditional distribution $P_l(X_2|X_1)$, and feeds back stably minimizable criticism of generated samples. The complete forward mapping generator can be learnt by minimizing losses (\ref{eq:fwd-gen}) and (\ref{eq:fwd-dis}) in an adversarial fashion \cite{NIPS2014_5ca3e9b1}. \looseness=-1
\vspace{-0.5mm}
\begin{flalign}
    & X_{12}  = G_2(E_1(X_1), c_2), ~c_2\sim \mathcal{N}(0,I) \label{eq:fwd-trans}\\
    & \mathcal{L}^{Tran}_{Gen} = \mathbb{E}_{X_1, c_2}~(D_2(X_{12}) - 1)^2\label{eq:fwd-gen}\\
    & \mathcal{L}^{Tran}_{Dis} =  \mathbb{E}_{X_2}~(D_2(X_{2}) - 1)^2 +\mathbb{E}_{X_1, c_2}~D_2(X_{12})^2 \label{eq:fwd-dis} \\
    & \mathcal{L}^{Tran}_{Pair} =  \mathbb{E}_{X_1, c_2, X_2} ~||~X^l_{12} - X^l_2~||_1 \label{eq:pair} \\
    &X_1 \sim P_l(X_1), ~X_2 \sim P(X_2)\nonumber
\end{flalign}
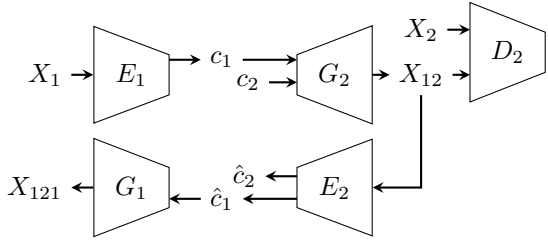
\begin{figure}[t!]
\centering

\begin{tikzpicture} [transform shape]

\tikzstyle{trapez} = [draw, trapezium, trapezium stretches=true, minimum height=1.0cm,
  text width=0.3cm, align=center]
  
\tikzstyle{arrow} = [thick,->,>=stealth]
  
\node (Encoder1) at (0,0) [trapez, rotate=270] {\rotatebox{-270}{$E_1$}};

\node (Decoder1) at (0,-1.5) [trapez, rotate=270] {\rotatebox{-270}{$G_1$}}; 

\node (Decoder2) at (2.7,0) [trapez, rotate=90] {\rotatebox{-90}{$G_2$}};

\node (Encoder2) at (2.7,-1.5) [trapez, rotate=90] {\rotatebox{-90}{$E_2$}};

\node (Discriminator) at (5.0,0.3
) [trapez, rotate=270] {\rotatebox{-270}{$D_2$}};

\draw node[left=0.8cm of Encoder1.center] (Template) {$X_1$};

\draw node[] at (3.85,0.0) (translatedSignal) {$X_{12}$};
\draw [arrow] (3.2,0) -- (translatedSignal);
\draw [arrow] (translatedSignal) -- (4.5, 0.0);

\draw node[] at (3.85,0.6) (trueSignal) {$X_{2}$};
\draw [arrow] (trueSignal) -- (4.5, 0.6);

\draw node[left= 0.8cm of Decoder1.center] (translatedSignalReturn) {$X_{121}$};
\draw node[right= of Encoder1.center, above] (encoder1Top) {};

\draw [arrow] (Template) -- (Encoder1);
\draw [arrow] (translatedSignal) |- (Encoder2);
\draw [arrow] (Decoder1) -- (translatedSignalReturn);


\draw node[] at (1.2,0.2) (contentCode1) {$c_{1}$};
\draw [arrow] (0.5,0.2) -- (contentCode1);
\draw [arrow] (contentCode1) -- (2.2, 0.2);

\draw node[] at (1.55,-0.1) (contentCode2) {$c_2$};
\draw [arrow] (contentCode2) -- (2.2, -0.1);

\draw node[] at (1.5,-1.35) (contentCodeTranslated2) {$\hat{c}_2$};
\draw [arrow] (2.2, -1.35) -- (contentCodeTranslated2);

\draw node[] at (1.2,-1.65) (contentCodeTranslated1) {$\hat{c}_1$};
\draw [arrow] (2.2, -1.65) -- (contentCodeTranslated1);
\draw [arrow] (contentCodeTranslated1) -- (0.5, -1.65);



\end{tikzpicture}
\caption{SilGAN - Translation stage} \label{fig:translation_stage}
\vspace{-5mm}
\end{figure}
The adversarial loss is normally sufficient to ensure both translation into a realistic maneuver and adherence to the profile in the template. However, we find that this adherence is sensitive to random seeding. To reliably ensure adherence while translating template $X_1$, a pairing loss (\ref{eq:pair}) is used to encourage reconstruction of the $l^{th}$ signal of the corresponding real maneuver $X_2$. Also, (\ref{eq:fwd-trans}) shows that upon encoding a template into $c_1$, $G_2$ can combine it with different codes $c_2$ to produce different maneuvers, all of which satisfy the profile requirements set by $c_1$. Such a technique of defining disentangled, partially shared, information domains for multi-modal domain translation, was first proposed in \cite{DBLP:conf/eccv/HuangLBK18}. \looseness=-1

Prior work has shown that the quality of domain translation can be improved by encouraging cycle-consistency \cite{DBLP:conf/iccv/ZhuPIE17}. That is, having translated template $X_1$ into a maneuver $X_{12}$, it is beneficial to reverse the translation and recover the template (\ref{eq:rev-trn}). We therefore define functions $E_2 : \mathcal{X}_{2} \rightarrow C_1 \times C_2$ and $G_1 : C_1 \rightarrow \mathcal{X}_1$ that reverse the translation. Unlike template extraction procedures used in preparing the training set which achieve the same end, reverse translation (\ref{eq:rev-trn}) is differentiable. This allows the cycle consistency objective (\ref{eq:cyc-cons}), the $L_1$ loss between input and recovered templates, to back propagate gradients from the reverse to the forward process during training. \looseness=-1
\begin{flalign}
    & X_{121} = G_1(\hat{c}_1), ~\hat{c}_1, \hat{c}_2 = E_2(X_{12}) \label{eq:rev-trn}\\
    & \mathcal{L}^{Tran}_{Cyc} = \mathbb{E}_{X_1, c_2} ~||~X_{121} - X_1~||_1 \label{eq:cyc-cons}
\end{flalign}
Forward and cycle-translation phases collectively constitute the translation stage (Figure \ref{fig:translation_stage}) of SilGAN. Networks in this stage use 1-D fully-convolutional layers, and can process templates of a duration of $N\leq512$, as long as $N$ is a power of 2. \looseness=-1

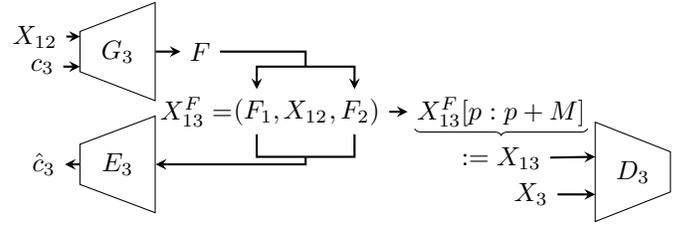
\begin{figure}[t!]
\centering

\begin{tikzpicture} [transform shape]

\tikzstyle{trapez} = [draw, trapezium, trapezium stretches=true, minimum height=1.0cm,
  text width=0.3cm, align=center]
  
\tikzstyle{square} = [rectangle, minimum width=1cm, minimum height=0.5cm,text centered, draw=black]
  
\tikzstyle{arrow} = [thick,->,>=stealth]
  
\node (Decoder3) at (0,0) [trapez, rotate=90] {\rotatebox{-90}{$G_3$}};

\node (Encoder3) at (0,-1.5) [trapez, rotate=90] {\rotatebox{-90}{$E_3$}};

\draw node[] at (1.05,-0.8) (fullExpSignal) {$X^F_{13} = $};
\draw node[] at (1.85,-0.8) (expSignal1) {$(F_1,$};
\draw node[] at (2.5,-0.8) (expSignal2) {$X_{12}$};
\draw node[] at (3.15,-0.8) (expSignal3) {$,F_2)$};

\draw node[] at (1.1,0) (expFull) {$F$};
\draw [arrow] (Decoder3) -- (expFull);
\draw [thick] (expFull) -| (2.5, -0.2);

\draw [arrow] (2.5, -0.2) -| (expSignal1);
\draw [arrow] (2.5, -0.2) -| (expSignal3);

\draw [thick] (expSignal1) |- (2.5, -1.4);
\draw [thick] (expSignal3) |- (2.5, -1.4);
\draw [arrow] (2.5, -1.4) |- (Encoder3);

\draw node[] at (5.1,-0.8) (expandedFull) {$X^F_{13}[p:p+M]$};
\draw [arrow] (expSignal3) -- (expandedFull);

\draw node[below= 0.4cm of expandedFull] (expandedNew) {};


\node (Disc3) at (6.85, -1.6) [trapez, rotate=270] {\rotatebox{90}{$D_3$}};

\draw[decoration={brace,mirror,raise=5pt},decorate] (3.95,-0.85) -- node[below=0.3cm of expandedFull] (newExpSign) {$:=X_{13}$} (6.25,-0.85);

\draw [arrow] (newExpSign) -- (6.35,-1.4);

\draw node[] at (5.5, -1.9) (trueSignal) {$X_{3}$};
\draw [arrow] (trueSignal) -- (6.35,-1.9);



\draw node[] at (-1.1, 0.2) (translatedSignal) {$X_{12}$};
\draw [arrow] (translatedSignal) -- (-0.5, 0.2);
\draw node[] at (-1.0, -0.2) (expansionCode) {$c_3$};
\draw [arrow] (expansionCode) -- (-0.5, -0.2);

\draw node[left=0.7cm of Encoder3.center] (expandedCode) {$\hat{c}_3$};
\draw [arrow] (Encoder3) -- (expandedCode);

\end{tikzpicture}
\caption{SilGAN - Expansion stage} \label{fig:expansion-stage}
\vspace{-3.5mm}
\end{figure}

\vspace{3mm}
\noindent \textbf{\textit{Expansion}} - One limitation of the translation stage is that it maps a template of duration $N$ into a maneuver of the same duration. Generating long maneuvers can therefore be cumbersome, since a template needs to be specified for its entire timeline. This is alleviated by defining an expansion stage $\mathcal{E}: \mathcal{X}_2 \rightarrow \mathcal{X}_3$ (Figure \ref{fig:expansion-stage}). Here, domain $\mathcal{X}_3$ comprises maneuvers $X_3 \sim P(X_3)$ of duration $M > N$. A common method for expanding a time series is forecasting. However, typical forecasting (backcasting) focuses only on the future (past) of a given series. The goal of expansion is different since it requires the generation of future and past transitions, both of which are jointly related to the given series. Here, $X_{12}$ is expanded into a longer maneuver $X_{13} \sim P_l(X_3 | X_1)$, by jointly generating plausible preceding ($F_1$) and succeeding ($F_2$) transitions for all $L$ signals. A learnable generator $G_3 : \mathcal{X}_2 \times C_3 \rightarrow \mathcal{F}_1 \times \mathcal{F}_2$, where the latent domain $C_3$ acts as the source of expanded information, generates extra parts (\ref{eq:exp-fill}) of length $(M-N)$ each. This ensures that cropping the concatenation $X^F_{13}$ according to (\ref{eq:exp-crop}) makes $X_{12}$ appear anywhere within $X_{13}$. Preceding and succeeding parts flexibly occupy the rest of the timeline. To ensure that expanded maneuvers are plausibly realistic, an additional least-squares discriminator $D_3 : \mathcal{X}_3 \rightarrow \mathbb{R}^+$ is defined. The function $G_3$ learns to produce realistic expansions in concert with $D_3$, by minimizing a novel \textit{random cropping adversarial loss}. This simple loss, defined by (\ref{eq:exp-gen}) and (\ref{eq:exp-dis}), operates by randomly cropping the concatenation $X^F_{13}$ using $p \sim \mathcal{U}\{0, M-N\}$ before evaluating it with $D_3$. This ensures that any crop of duration $M$ is realistic. \looseness=-1
\begin{flalign}
    & X^F_{13} = (F_1, X_{12}, F_2), ~F_1, F_2 = G_3(X_{12}, c_3)\label{eq:exp-fill}\\
    & X_{13} = X^F_{13}[p: p+M] \label{eq:exp-crop}\\
    & \mathcal{L}^{Exp}_{Gen} = \mathbb{E}_{X_1, c_2, c_3,p}~(D_3(X_{13}) - 1)^2\label{eq:exp-gen}\\
    &\mathcal{L}^{Exp}_{Dis} = \mathbb{E}_{X_3}~(D_3(X_{3}) - 1)^2 + \mathbb{E}_{X_1, c_2, c_3, p}~D_3(X_{13})^2
    \label{eq:exp-dis}\\
    & X_3 \sim P(X_3), ~ c_2, c_3 \sim \mathcal{N}(0,I), ~p \sim \mathcal{U}\{0, M-N\} \nonumber
\end{flalign}
To showcase expansion, we fix $M=1024$ so that any template of duration $N \leq 512$, as long as it is a power of two, can be expanded to a maneuver of duration 1024. To train the expansion stage, a separate set of recorded maneuvers of this duration is collected. While networks in the translation stage use 1-D convolution, 2-D convolution is found to be more effective in expansion. With an increased duration, stacking up parts of the timeline and applying a 2-D kernel seems more effective. \looseness=-1

\vspace{3mm}
\noindent \textbf{\textit{Bidirectional reconstruction}} - Upon observing the layout of either stage, one can note the presence of autoencoders $E_i((G_i)), i=1, 2, 3$. The learning process can therefore be eased using objectives (\ref{eq:id-rec}), which encourage them to learn identity mapping in the respective domains. This, however, does not apply to the autoencoder in the expansion stage, since it does not operate directly on the domain $\mathcal{X}_3$. In addition, as shown in \cite{DBLP:conf/eccv/HuangLBK18}, learning identity mappings in the code domains using objectives (\ref{eq:code-rec1}) and (\ref{eq:code-rec2}), is a simple way of encouraging code distributions to match $\mathcal{N}(0,I)$. \looseness=-1
\begin{flalign}
    & \mathcal{L}_{ID_i} = \mathbb{E}_{X_i} ~||~X_i - G_i(E_i(X_i))~||_1, i=1, 2
    \label{eq:id-rec}\\
    & \mathcal{L}_{CR_2} = \mathbb{E}_{X_1, c_2} ~||~(c_1, c_2) - E_2(G_2(c_1, c_2))~||_1 \label{eq:code-rec1}\\
    & \mathcal{L}_{CR_3} = \mathbb{E}_{X_1, c_2, c_3} ~||~c_3 - E_3(G_3(X_{12}, c_3))~||_1 \label{eq:code-rec2}\\
    &c_1 = E_1(X^l_1),~ c_2, c_3 \sim \mathcal{N}(0,I)\nonumber
\end{flalign}
This is not only important to ease sampling, but also essential for generalizing the models generative capabilities. Such a simple code reconstruction objective avoids complex alternatives like Kullback-Leibler loss, while also, as empirically observed, producing translations of a better quality.

\noindent \textbf{\textit{End-to-end training}} - Combining all training objectives previously defined, SilGAN is trained end-to-end by minimizing the composite adversarial objective (\ref{eq:comp-obj}).
\vspace{-1mm}
\begin{flalign}
    &\mathcal{L}_{Gen} = \min_{E_1, E_2, G_1, G_2} \lambda_{Gen}\big(\mathcal{L}^{Tran}_{Gen} +\mathcal{L}^{Exp}_{Gen}\big) + \lambda_{Pair} \mathcal{L}^{Tran}_{Pair}
    \nonumber\\
    &+ \lambda_{Cyc}\mathcal{L}^{Tran}_{Cyc} + \lambda_{ID} \sum_{i=1,2} \mathcal{L}_{ID_i} + \sum_{j=2,3} \lambda_{CR} \mathcal{L}_{CR_j} \nonumber\\
    &\mathcal{L}_{Dis} = \min_{D_2, D_3} \big(\mathcal{L}^{Tran}_{Dis} +\mathcal{L}^{Exp}_{Dis}\big)\nonumber \\
    &\mathcal{L} = \mathcal{L}_{Gen} + \mathcal{L}_{Dis}\label{eq:comp-obj}
\end{flalign}
Results shown here are sampled from a model trained with relative importance for all objectives set to 1, except $\lambda_{ID}=10$. Further details about training are available in the released code. Depending upon the configuration, training the model takes 5--15 minutes per epoch on a single NVidia Tesla V100 GPU. To evaluate the quality of generation, a held-out test set of $\sim$20$k$ recorded maneuvers, each of duration 512$s$, and corresponding extracted templates is used. As an end-to-end indication of the quality of translation, cycle reconstruction of the test set is measured using (\ref{eq:cyc-cons}). However, instead of $L_1$ distance, the easier to interpret Structural Similarity Index (SSIM) is used. SSIM between two time series is a real number in $[0, 1]$, with $0$ indicating dissimilarity and $1$ indicating a perfect match. Cycle reconstruction is measured as the average SSIM between a test template $X_1$ and 4 cycle translations $X_{121}$, each using a different random code $c_2$. After training for $\sim$15 epochs, cycle reconstruction SSIM, averaged over the entire test set, settles around $0.95$. While such high similarity of cycle reconstruction indicates healthy training, quality of forward translation is additionally ensured by visually inspecting around hundred randomly selected test template translations. Visual inspection of time series maneuvers confirms adherence to the template and plausibility of characteristic features like correlations between vehicle and engine speed under the influence of gear shifts.  \looseness=-1
\begin{figure}[t]
    \centering
    \includegraphics[width=0.9\linewidth]{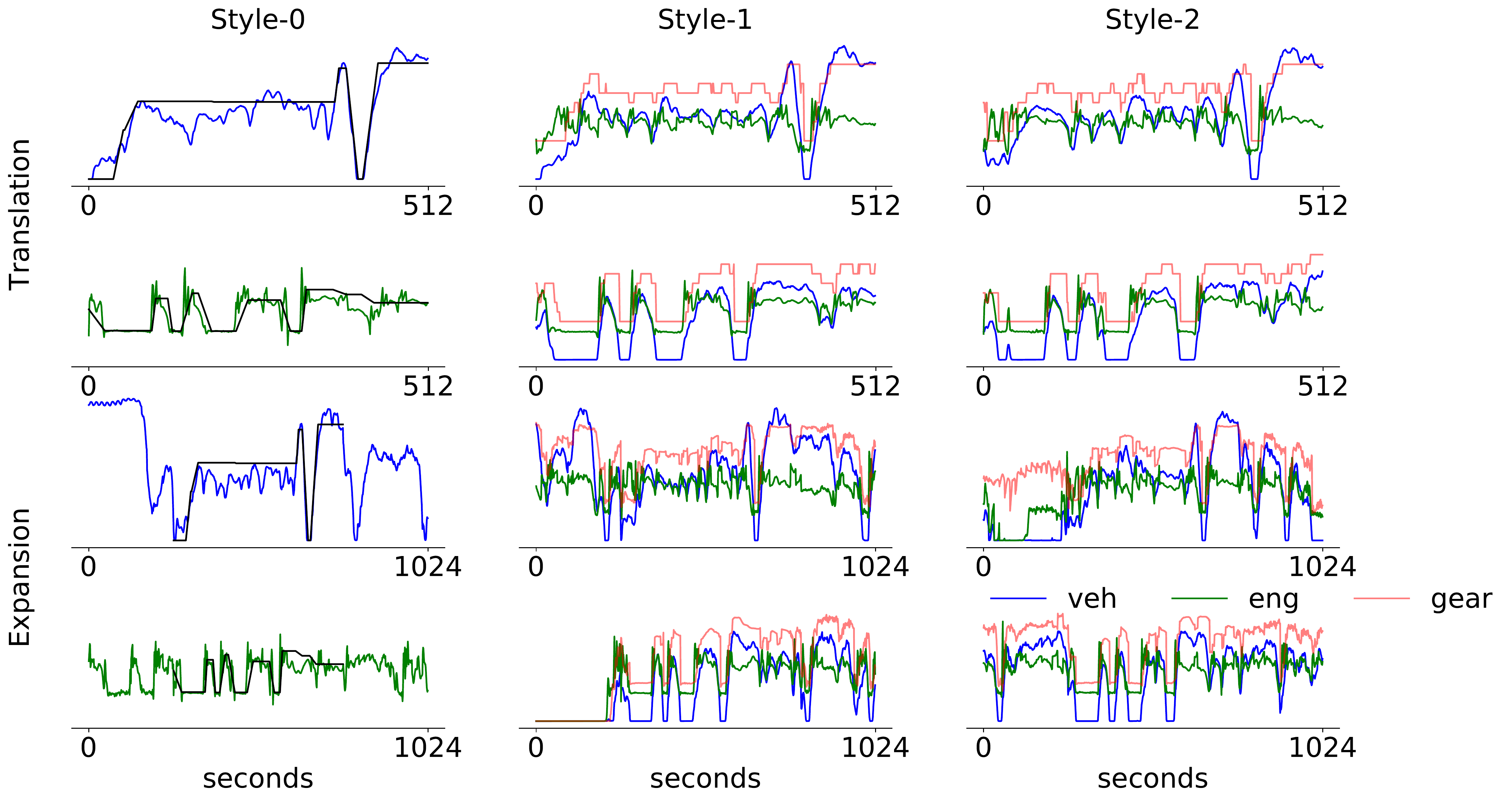}
    \caption{Generation of maneuvers by specifying one vehicle speed and one engine speed template. The first column shows the template overlaid on the corresponding generated signal.\looseness=-1}
    \label{fig:single-scenario}
    \vspace{-6mm}
\end{figure}

\section{Using SilGAN for test stimulus generation}
\label{sec:gen}
Upon training, simple scenario-based generation of realistic driving maneuvers is accomplished by specifying a single template. As shown in Figure \ref{fig:single-scenario}, a template specified for any one constituent signal can be translated and expanded into diverse, but realistic, interpretations. Here, the choice $p=\frac{M-N}{2}$ ensures that the translated maneuver appears at the center of the expanded timeline. It can be seen that translations are quite faithful in adhering to the template, while still showing subtle variations in its details, across interpretations. The expanded maneuvers, on the other hand, show substantial diversity in preceding and succeeding signal transitions.

\begin{figure}[t]
    \centering
    \includegraphics[width=0.9\linewidth]{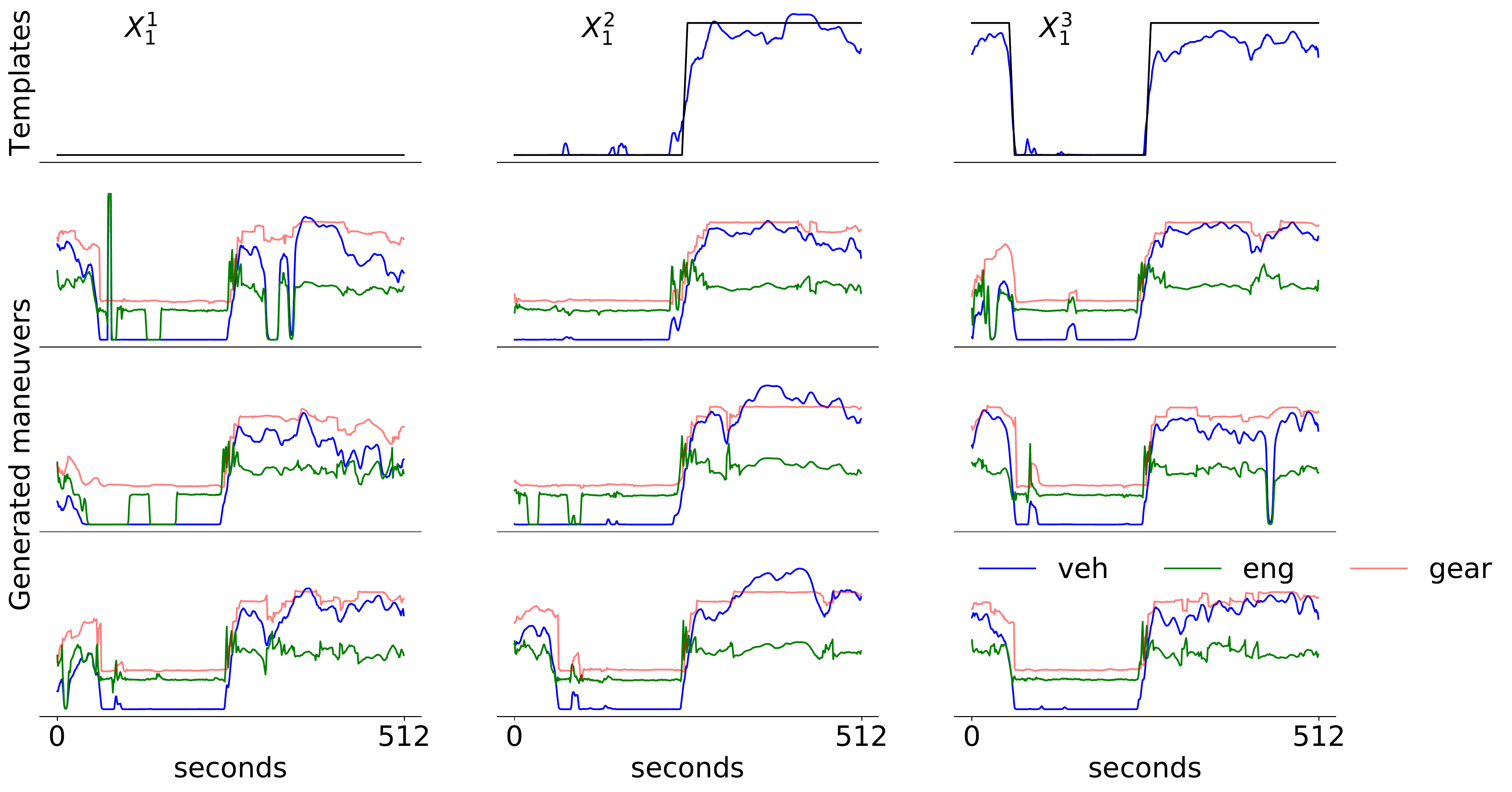}
    \caption{By specifying three templates (top row) -- null, takeoff and stop before takeoff, generating maneuvers using (\ref{eq:simplex}) are constrained to combine characteristics of all three}
    \label{fig:combined-takeoff-sample}
    \vspace{-8mm}
\end{figure}

A more potent way of using SilGAN would be with multiple templates, ensuring compliance with a combination of scenarios. It is well-known \cite{DBLP:journals/corr/RadfordMC15} that generative models achieve nonlinear combination of characteristics in the sample space using trivial linear combinations in its latent space. As noted earlier, systematically generating a variety of takeoff maneuvers is valuable for testing critical functions, and Figure \ref{fig:combined-takeoff-sample} shows how this can be easily achieved. Let us start by specifying a helpful \textit{null} vehicle speed template, $X^1_1$, where the vehicle is completely still for 512 seconds. Let us then specify one takeoff template $X^2_1$ where the vehicle stays still for half that time before beginning to roll. Since this may be considered too narrow a takeoff scenario, let us widen it by specifying an additional template $X^3_1$, where the vehicle stops once before taking off. The triple of codes $c^{(3)}_1 = (E_1(X^1_1), E_1(X^2_1), E_1(X^3_1))^T$ can be imagined as the vertices of a triangle, with each vertex being embedded in the high-dimensional latent domain $C_1$. Sampling any point on this triangle, along with a random code $c_2$, and decoding them using $G_2$, guarantees that generated maneuvers do not stray beyond the driving conditions collectively posed by the three chosen templates (Figure \ref{fig:combined-takeoff-sample}). Such constrained, but realistic, generation produces authentic stimuli for SIL testing under a controlled composite test scenario. This extends to the general case of using $K$ templates, meaning that it is possible to generate maneuvers that is guaranteed to comply with a scenario of an arbitrary level of sophistication. Sampling in the latent hyper-plane with $K$ vertices is achieved by drawing a simplex $\alpha^{(K)}$, $\sum^K_{i=1} \alpha^{(K)}_i = 1$, from a Dirichlet distribution of order $K$, and linearly combining it with codes $c^{(K)}_1$ (\ref{eq:simplex}). Repeatedly drawing simplexes also ensures that the hyper-plane is uniformly covered, making it an apt tool for exploring the variety of combinations that are possible. \looseness=-1
\begin{flalign}
    &\alpha^{(K)} \sim Dir(K, \mathbbm{1}^K), ~c^{(K)}_1 = \big(E_1(X^1_1), ..., E_1(X^K_1)\big)^T \nonumber\\
    & X_{12} = G_2(\alpha^{(K)}\cdot c^{(K)}_1, c_2), ~c_2 \sim \mathcal{N}(0,I) \label{eq:simplex}
\end{flalign}
\section{Using SilGAN for test automation}
\label{sec:aut}
Let us choose templates $X^{(3)}_1 =  (X^1_1, X^2_1, X^3_1)^T$, described previously, as a test scenario for takeoff. In the software under test, let us assume there is function $agg\_to$ (Figure \ref{fig:code-examples}a) that observes vehicle and engine speed signals to check whether the engine largely idles before the vehicle quickly takes off and cruises at increasing speeds. One can find such a function as part of logic that identifies aggressive driving events. Events, so identified, are either logged for analysis or are actively compensated, for example, by adjusting fuel consumption or exhaust after treatment. Under the takeoff test scenario $X^{(3)}_1$, let the objective be to find a maneuver $\mathcal{H}$ that checks the design of takeoff defined by $agg\_to$. This can be seen as a form of code coverage, i.e. finding a stimulus that satisfies the assumptions in the \textit{if} condition that appears in the test function. While simplex sampling using (\ref{eq:simplex}) may be effective for generating a variety of maneuvers that comply with the defined scenario, the number of possibilities is infinite. Finding $\mathcal{H}$ by random sampling is therefore an impractical search technique. More importantly, $agg\_to$ indicates coverage solely by returning 1 when the \textit{if} condition is satisfied. It therefore poses a fundamental problem in providing no real-time feedback to guide the sampling process. \looseness=-1

\begin{figure}[t]
    \centering
    \begin{minipage}[b]{.4\linewidth}
\begin{lstlisting}[language=Python]
def agg_to(v, e):
 m1 = mean(v[256:350])
 m2 = mean(v[350:512])
 m3 = mean(e[0:256])
 if(0.49 < m1 < 0.51) & 
   (0.59 < m2 < 0.61) & 
   (m3 < 0.08):
  ret1 = 1
 else:
  ret1 = 0
 return ret1
 
 
 
 
 
\end{lstlisting}    
        \caption*{(a)}
        \label{fig:test-code}
    \end{minipage}\qquad
    \begin{minipage}[b]{.4\linewidth}
\begin{lstlisting}[language=Python]
def S(v, e):
  m1 = mean(v[256:350])
  m2 = mean(v[350:512])
  m3 = mean(e[0:256])
  c0 = and(lt(0.49, m1), 
     lt(m1, 0.51))
  c1 = and(lt(0.59, m2), 
     lt(m2, 0.61))
  c2 = lt(m3, 0.08)
  c = and(and(c0, c1), c2)
  return c
\end{lstlisting}    
        \caption*{(b)}
        \label{fig:search-code}
    \end{minipage}
    \caption{A simple function (a) in software under test which is automatically transformed into a search function (b) that returns one coverage indicator per branching condition} 
    \label{fig:code-examples}
    \vspace{-6mm}
\end{figure}

Code coverage entails satisfying branching conditions, each of which is typically defined using a composition of boolean operations. Since these operations evaluate to only \textit{true} or \textit{false}, there is no indication of how distant a test input is from making an \textit{if} condition evaluate to, say, \textit{true}. However, cases like the chosen test function, where branching conditions evaluate real values, are fairly routine in decision making logic found in vehicle control software. In such cases, it is possible to convert discrete boolean operations into real-valued measures. Setting aside boolean operators $=$ and $\neq$ which are not directly applicable to real values, atomic operations $<$ and $>$ can be readily converted into difference functions (\ref{eq:bool-diff}). Arguments that evaluate to a large positive value indicate significant mismatch, and those that evaluate to a negative value indicate a certain match for both conditions. This also means that $\neg$, i.e. guaranteeing the mismatch of an atomic condition, is simply achieved by arithmetic negation\footnote{Defining $not$ as arithmetic negation restricts it to a subset of possible solutions. For example $\neg(a < b) = a \geq b$, but $not$ only finds solutions $a>b$, setting aside equality checks on real values}. This principle extends to $\land$ and $\lor$, where functions $max$ and $min$ respectively check whether all, or any, atomic operation(s) evaluate to a negative value. While alternative definitions are certainly possible, chosen real-valued equivalents of boolean operations (\ref{eq:bool-comp}) are not only simple but also differentiable\footnote{$max(c,d)$ and $min(c,d)$ are not differentiable at $c=d$, but this can be disregarded for real-valued $c$ and $d$}. One simple check of consistency is that these functions satisfy De Morgan's laws\footnote{$not(and(a,b)) = -max(a,b) = min(-a,-b) = or(not(a), not(b))$}. \looseness=-1
\vspace{-1mm}
\begin{flalign}
    &\textbf{\textit{lt}}(a, b) = a-b ~~~~~~~~~~~~~\textbf{\textit{gt}}(a,b) = b-a \label{eq:bool-diff}\\
    &\textbf{\textit{and}}(c,d) = max(c,d) ,~~~\textbf{\textit{or}}(c,d) = min(c,d) \nonumber\\
    &\textbf{\textit{not}}(c) = -c \label{eq:bool-comp}
\end{flalign}
\vspace{-5mm}
\begin{figure}[t]
    \begin{center}
    \begin{tikzpicture}
        \tikzstyle{arrow} = [thick,->,>=stealth]
        \node (loss) {\includegraphics[width=0.7\linewidth]{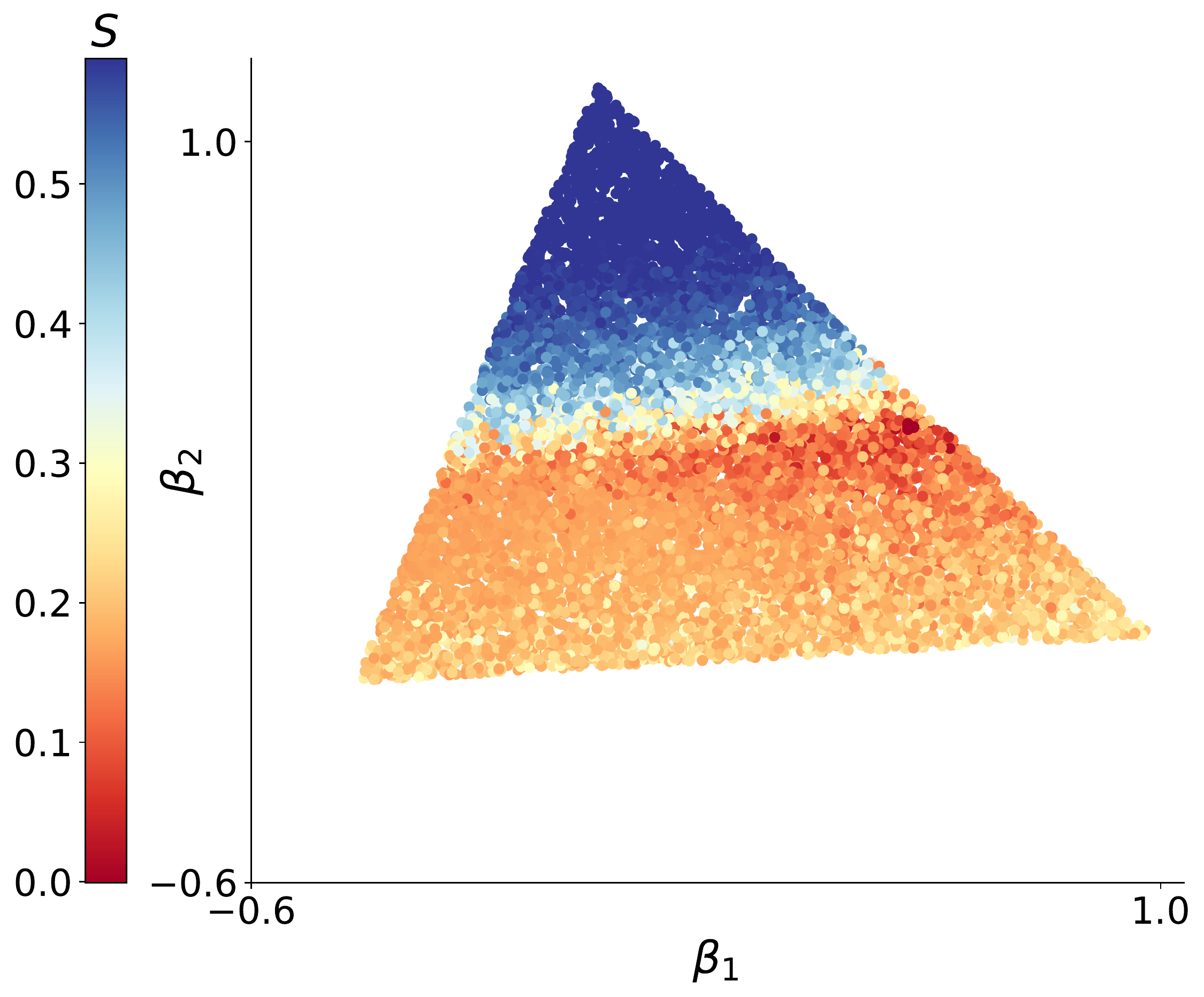}};
        \node (gd_reg) at(loss.center)[draw, black, line width=2pt, ellipse, minimum width=5pt, minimum height=15pt, rotate=-28, xshift=35pt, yshift=30pt, rotate=-50]{};
        \node (hit) at (2.3, -1.35) {\includegraphics[width=0.25\linewidth]{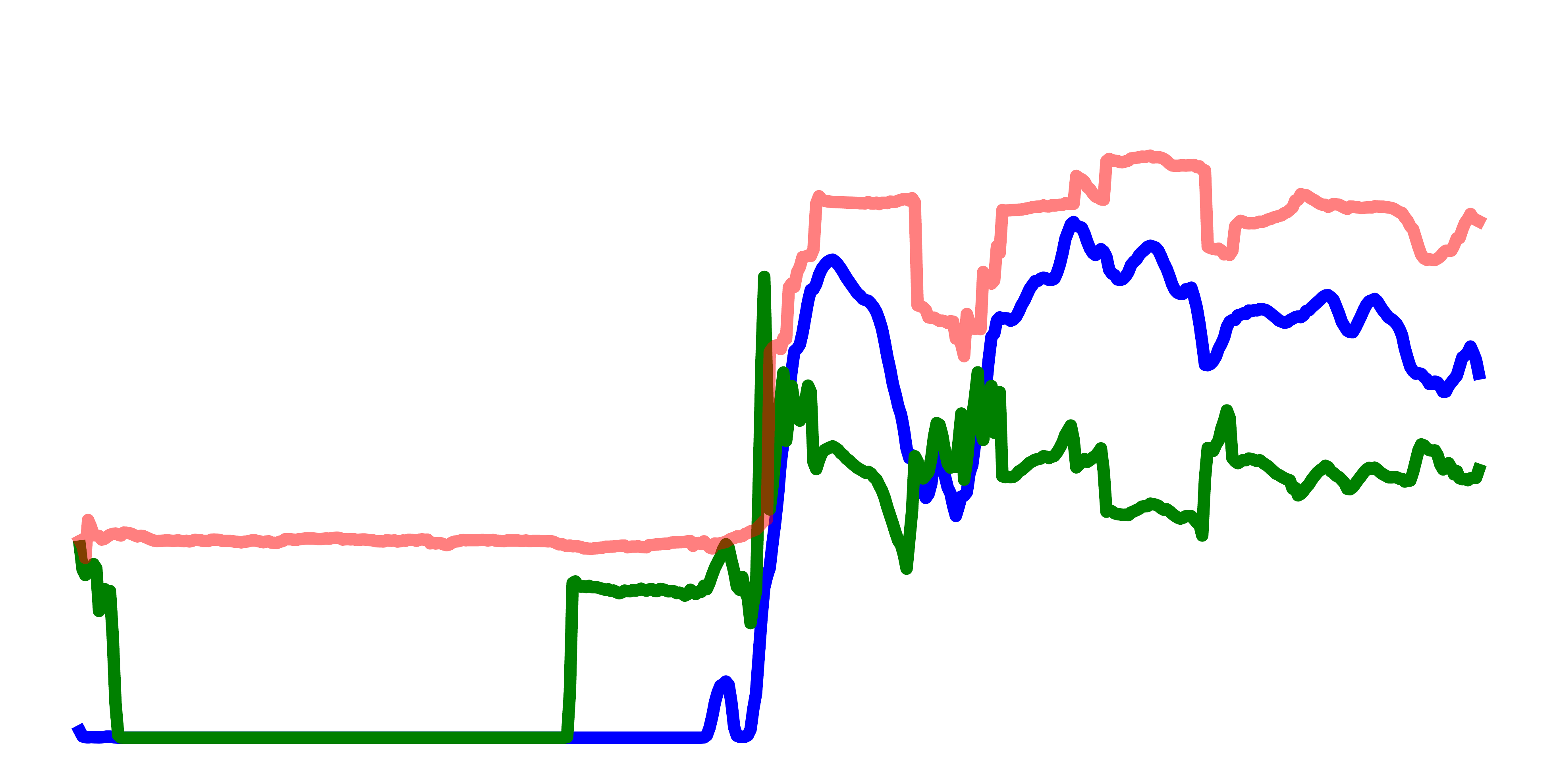}};
        \draw [arrow] (1.58, 0.42) -- (2.1, -1.3);
        \node[align=left] at (0.4, -0.3) {\textbf{Simplex sampling}};
        \node[align=left] at (3.0, 0.9) {\textbf{Gradient descent}};
        \node[align=left] at (1.5, -1.1) {\textbf{$\mathcal{H}$}};
        \node[align=left] at (-0.3, 2.1) {\large \textbf{$c^1_1$}};
        \node[align=left] at (3.2, -0.6) {\large \textbf{$c^2_1$}};
        \node[align=left] at (-1.5, -0.9) {\large \textbf{$c^3_1$}};
        
        
    \end{tikzpicture}
    \end{center}
    \vspace{-5mm}
    \caption{With test scenario $X^{(3)}_1$, automatic code coverage for the test function in Figure \ref{fig:code-examples} begins with simplex sampling in the triangle with vertices $c^{(3)}_1$ in the latent space $C_1$. Once a promising area is found, gradient-descent takes over to find $\mathcal{H}$ that satisfies the condition. For visualization in 2-d, we plot the triangle by taking steps $\beta_1,\beta_2$ along two randomly chosen basis vectors that span its plane. A similar search is jointly conducted in $C_2$, but it is not visualized here.\looseness=-1} 
    \label{fig:loss-landscape}
    \vspace{-3mm}
\end{figure}

In transforming boolean operations on real values into another real value, these functions give continuous feedback on how distant a test input is from covering a branching condition on real values. We therefore refer to them as \textit{coverage indicators}. By traversing its abstract syntax tree, substituting boolean operators with coverage indicators, and extracting a composite indicator for its \textit{if} condition, the chosen test function is automatically transformed into a search function $S : \mathcal{X}_2 \rightarrow \mathbb{R}$ (Figure \ref{fig:code-examples}b). This means $S<0$ guarantees that the corresponding input maneuver is a $\mathcal{H}$ that satisfies the branching condition. Further, since the coverage indicators are readily differentiable, as long as operations leading up to branching conditions are also differentiable, the resulting $S$ is itself differentiable. This means that the right combination of $\alpha^{(N)}$ and $c_2$, that generates $\mathcal{H}$, can be found using gradient descent. \looseness=-1

While the search function may be differentiable, its multi-dimensional landscape is likely to feature many local minima. As seen in Figure \ref{fig:loss-landscape}, the presence of large flat regions, where $S$ hardly changes, clearly hinders search by gradient descent. To address this issue, we propose a novel search method that \textit{combines sampling and gradient-descent} to automatically generate test stimuli that match the design of code under test (Algorithm \ref{alg:sil-coverage}). This method begins with simplex sampling for $n\_sim$ iterations in order to survey the coverage landscape in the triangle $c^{(3)}_1$. If the sampling process chances upon $\mathcal{H}$, the objective is achieved, ending the search. Otherwise, the algorithm identifies a promising area, i.e. a combination of $\alpha^*$ and $c^*_2$ that generates a maneuver resulting in $S \approx 0$, around which finding a $\mathcal{H}$ is likely. The search is then taken over by gradient descent, which attempts to iteratively minimize $S$ by jointly varying $\alpha^*$ and $c^*_2$ along directions of fastest decrease in $S$. Reparameterizing $\alpha^*$ using logit/sigmoid ensures that gradient-descent does not stray outside the triangle, guaranteeing compliance with the test scenario. This search persists for a maximum of $n\_gd$ steps during which, if $S$ becomes negative, $\mathcal{H}$ has been found and the search is complete. Otherwise, the search times out without finding $\mathcal{H}$ under the given test scenario. \looseness=-1

\begin{algorithm}
\caption{Scenario-based test automation}
\label{alg:sil-coverage}
\SetKwInOut{Input}{Input}
\SetKwInOut{Output}{Output}
\SetKwInOut{Parameters}{Parameters}
\begin{small}
    \Input{Scenario of templates $X^{(K)}_1$, search function $S$\looseness=-1}
    \Output{Maneuver $\mathcal{H}$ satisfying the coverage condition}
    \Parameters{$n\_sim$ - max sampling steps, $n\_gd$ - max gradient descent steps, $\eta$ - gradient step size}
    $c^{(K)}_1 = \big(E_1(X^0_1), ..., E_1(X^K_1)\big)^T$\\
    $s_{min} = \infty$,~ $\alpha^* = null$, ~$c^*_2=null$\\
    \For {$n\in[n\_sim]$}{
        $\alpha^{(K)} \sim Dir(K, \mathbbm{1}^K)$, $c_2 \sim \mathcal{N}(0,I)$  \\
        $\mathcal{H} = G_2(\alpha^{(K)} \cdot c^{(K)}_1, c_2)$ \\
        \uIf{$S(\mathcal{H})<0$}{
            \Return $\mathcal{H}$
        }
        \ElseIf{$S(\mathcal{H})<s_{min}$}{
            $\alpha^* = \alpha^{(K)}$, ~ $c^*_2 = c_2$, ~  $s_{min}=S(\mathcal{H})$
        }
    }
    $\gamma^* = \textit{logit}(\alpha^*)$\\
    \For {$n\in[n\_gd]$}{
        $\alpha^* = sigmoid(\gamma^*)$\\
        $\alpha^* = \alpha^*/sum(\alpha^*)$\\
        $\mathcal{H} = G_2(\alpha^* \cdot c_1, c^*_2)$ \\
        \uIf{$S(\mathcal{H})<0$}{
            \Return $\mathcal{H}$
        }
        $\gamma^* = \gamma^* - \eta \nabla_{\gamma^*} (S)(\mathcal{H})$\\
        $c^*_2 = c^*_2 - \eta \nabla_{c^*_2} (S)(\mathcal{H})$\\
    }
\end{small}
\end{algorithm}

Figure \ref{fig:loss-landscape} shows the case of simplex sampling being unable to find $\mathcal{H}$ even after setting $n\_sim\approx100k$. Combining the respective strengths of sampling and gradient-descent, both of which use feedback from coverage indicators, it is often sufficient to sample for 10--50 iterations, with the subsequent gradient search taking no more than a few 10s of iterations to find $\mathcal{H}$. The example in Figure \ref{fig:code-examples} may contain a single branching condition, but the method can be extended to multiple independent \textit{if-else} conditions, even when they are nested. A search function, in this case, returns a vector containing one minimizable coverage indicator per independent branching condition. Due to independence, this vector can be collectively minimized by executing one search per coverage indicator in parallel. \looseness=-1

Using SilGAN, we thus demonstrate an end-to-end pipeline for scenario specification, stimulus generation and test automation. In demonstrating test automation, we however show only a case of code coverage with boolean operations made differentiable using coverage indicators. While the principle of targeted, combined sampling and gradient based search of the latent space is vital, additional measures may be necessary to scale the technique for automatically testing code with non-differentiable intermediate operations. We leave the investigation of such measures for future work. \looseness=-1

\section{Related work}
Deep generative models for time series have garnered a substantial amount of attention in a variety of applications. Examples include generating synthetic medical data to ensure privacy of the patients \cite{DBLP:journals/corr/EstebanHR17}, modelling financial time series to forecast asset returns \cite{DBLP:journals/corr/abs-1907-06673} and music creation \cite{DBLP:journals/corr/Mogren16}. In contrast, domain adaptation of time series has received sparser attention. Previous work has employed domain adaptation to extract physiologically invariant features in a clinical setting \cite{DBLP:journals/corr/abs-1904-00655} and the creation of a dataset invariant to specific characteristics of individual blast furnaces \cite{DBLP:journals/corr/abs-2007-07518}. Closest to our work is \cite{DBLP:conf/aitest/ParthasarathyBH20} that seeks to generate in-vehicle time series using a weak form of domain adaptation. We significantly improve on that work by introducing a simpler way of using templates and using it to set explicit objectives for domain adaptation. This greatly reduces specification effort and eases generation. Moreover, we enrich test stimulus generation through multi-modal translation using disentangled information domains and expansion. The technique for targeted search further makes it much more conducive for SIL testing. While parallels to our overall approach are rare in the domain of time series, certain aspects have analogies in the image domain. Examples include sketch to image \cite{DBLP:conf/cvpr/IsolaZZE17}, \cite{DBLP:conf/iccv/ZhuPIE17}, \cite{DBLP:conf/eccv/HuangLBK18}, and image outpainting \cite{DBLP:journals/corr/abs-1808-08483}. \looseness=-1

Prior research on embedding methods for time series include \cite{DBLP:conf/eann/NalmpantisV19} that trains general purpose embeddings through tokenization and skip-gram techniques. By exploiting the fact that similarity matrices can be of low rank, \cite{DBLP:conf/ijcai/LeiYVWD19} converts time series into a matrix representation suitable for clustering techniques. In \cite{DBLP:journals/corr/abs-1906-05205}, a twin-autoencoding structure is employed to learn embeddings invariant to warped transformation. Our technique for embedding differ by disentangling time series at different levels of abstractions into components with distinct interpretations, apt for a testing pipeline.\looseness=-1 

Previous work on deep learning for test case generation include time series generation for automotive SIL testing \cite{DBLP:conf/aitest/ParthasarathyBH20}, text generation for testing mobile apps \cite{DBLP:conf/icse/0010ZPZMZ17}, and protocol frame generation for testing process control equipment \cite{DBLP:conf/icst/ZhaoLWSH19}. Parallel work on dynamic software testing, include identifying worst case branching for stress testing \cite{DBLP:conf/icst/KooS0B19} and GUI testing for apps \cite{DBLP:conf/issta/PanHWZL20}. Unlike these methods which target specific aspects of testing in different domains, by introducing techniques for specification, generation, and test automation, we demonstrate an end-to-end framework for SIL testing of vehicle control software. \looseness=-1

Finally, while we show a simple example of using the trained model for SIL test automation, in extending our work to more general cases, previous work on ways to smooth boolean conditions \cite{DBLP:journals/corr/abs-1802-04408} \cite{DBLP:journals/tcad/EddelandCSRMA20} may be helpful. \looseness=-1
\section{Conclusions}
\label{sec:conc}
The future of automotive software testing depends upon techniques that accelerate feedback without compromising quality. Exploiting the ready availability of data from field vehicles and advances in deep learning, we present a domain translation model that helps achieve both. Using this model, we demonstrate how to specify and generate realistic driving maneuvers at a granular level, easing the test of low-level vehicle control software. Further, we also introduce techniques of targeted generation that can automate test objectives like code coverage. While improvements such as scaling the automation procedure to deal with other test objectives can strengthen the method, its fundamental principles significantly increase the credibility of simulation-driven testing as a quicker and viable alternative to field tests. \looseness=-1

\section{Acknowledgements}
We thank Carl Seger and Henrik Lönn for helpful discussions. This work is supported by the Wallenberg Artificial Intelligence, Autonomous Systems and Software Program (WASP) funded by the Knut and Alice Wallenberg Foundation. \looseness=-1

\bibliographystyle{ieeetr}
\bibliography{bibliography}

\end{document}